\documentclass[aps,prb,numerical,a4paper,showpacs,twocolumn]{revtex4-1}

\usepackage[dvipdfmx]{graphicx}%
\usepackage[dvipdfmx]{color}%
\usepackage{dcolumn}%
\usepackage{bm}%
\draft 

\begin{document}


\title{Inductive intrinsic localized modes in a 1D nonlinear electric transmission line} 



\author{M. Sato}
\email[]{msato153@staff.kanazawa-u.ac.jp}
\author{T. Mukaide}
\author{T. Nakaguchi}
\affiliation{Graduate School of Natural Science and Technology, Kanazawa University\\Kanazawa, Ishikawa 920-1192, Japan}
\author{A. J. Sievers}
\affiliation{Laboratory of Atomic and Solid State Physics, Cornell University\\
Ithaca, NY 14853-2501, USA}

\date{\today}

\begin{abstract}
The experimental properties of intrinsic localized modes (ILM) have long been compared with theoretical dynamical lattice models that make use of nonlinear onsite and/or nearest neighbor intersite potentials. Here it is shown for a 1-D lumped electrical transmission line  a nonlinear inductive component in an otherwise linear parallel capacitor lattice makes possible a new kind of ILM outside the plane wave spectrum. To simplify the analysis the nonlinear inductive current equations are transformed to flux transmission line equations with analogue onsite hard potential nonlinearities. Approximate analytic results compare favorably with those obtained from a driven damped lattice model and with eigenvalue simulations. For this mono-element lattice ILMs above the top of the plane wave spectrum are the result. We find that the current ILM is spatially compressed relative to the corresponding flux ILM. Finally this study makes the connection between the dynamics of mass and force constant defects in the harmonic lattice and ILMs in a strongly anharmonic lattice.
\end{abstract}


\maketitle 


\section{Introduction}
An intrinsic localized mode (ILM)\cite{1}, often referred to as a discrete breather (DB)\cite{2,3}, is a characteristic localized vibrational excitation in a periodic lattice with nonlinear potential energy. The energy profile of a stationary ILM resembles that of a force constant defect in a harmonic lattice\cite{4,5} but like a soliton, it can propagate; however, in contrast to a soliton it looses energy as it moves through the lattice. The theoretical, numerical and experimental properties of these localized excitations have been summarized in a number of reviews, often focusing on the different kinds of applications: they range from micro-nanomechanical\cite{6}, to superconducting\cite{7}, magnetic\cite{8}, optical\cite{3}, lattice dynamical\cite{9,10} and defect formation\cite{11,12}. 

In the lattice dynamical studies of ILMs nonlinearity enters the dynamics through the nonlinear properties of the effective intersite and/or onsite potentials and the inertial component is strictly linear. In other fields it has been recognized that nonlinear inertial contributions do occur. The large amplitude, strongly nonseparable, collective motion in the vibration-rotation dynamics of nuclei represents such a case\cite{13,14,15}. The coordinate dependent vibrational and rotational masses that produce high precision energy levels for the spectrum of the ${\rm H}_3^ +  $  molecule characterize yet a different class\cite{16}. These different demonstrations have encouraged us to consider the dynamical possibilities of a new type of ILM in a 1-D nonlinear transmission line. The dynamical properties of amplitude dependent inertial masses for strongly non-separable modes in a nonlinear vibrational lattice have not yet been treated; however, nonlinear lumped element electrical transmission line studies have a long history\cite{17,18,19} and there is a well known translation between inertial mass and electrical inductance for such linear transmission lines\cite{20}. As long as electrical pulses extend over many nonlinear elements of an electrical transmission line so that continuum equations, such as the Korteweg-de Vries, could be applied it has been possible to make contact with soliton behavior\cite{23,21,22,24,25}. In more recent times interest has shifted from understanding soliton behavior to the production of high frequency radiation using electromagnetic shock waves produced by hysteresis in nonlinear electric lines\cite{26,27,28}. Fundamental studies focusing on a localized nonlinear excitation with width comparable to the lattice constant of a lumped electrical array have appeared in the last decade\cite{29,30,31,32}. To date all of these ILM systems have made use of nonlinear capacitors to produce intersite nonlinear coupling between the linear inductor lattice sites.

In this report we describe a different kind of ILM associated with nonlinear inductors equally spaced in an otherwise linear electrical transmission line. A 1-D electric lattice with linear intersite capacitance coupling plus current dependent inductance (without hysteresis) is the starting point for the development of such an ILM and its production is studied using three different methods: approximate analytic, driven-damped and eigenvector simulations. All three methods are in good agreement and show that the current ILM is more focused than the corresponding flux ILM and that for the limit of large driving amplitude the flux ILM excitation approaches localization to three cells while the corresponding current ILM reaches a single lattice cell excitation.

\begin{figure}
\includegraphics{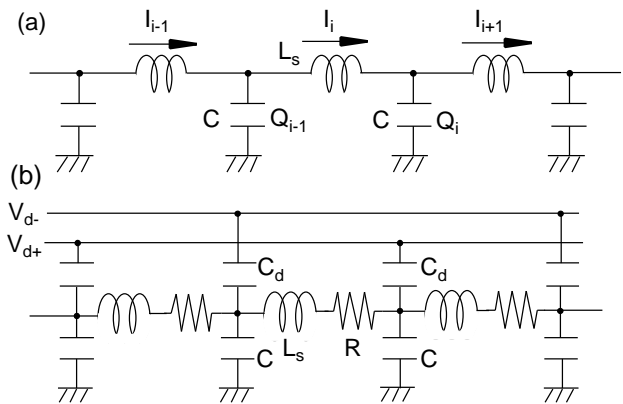}%
\caption{\label{fig:1}
(a) Circuit diagram for the electric transmission line with a saturable nonlinear inductor  $L_s$ and linear capacitor $C$.
(b) Circuit for the driven-damped system. Resistance, $R$. The line is driven by an oscillator via $V_{d+}$  and $V_{d-}$  through coupling capacitor $C_d$.
}
\end{figure}

\section{THREE SOLUTIONS TO THE FLUX EQUATIONS OF MOTION}
\subsection{Approximate analytic method}
The transmission line under consideration is shown in Fig.~\ref{fig:1}(a). The array consists of linear capacitors C connected to coils; each of $n$ turns rapped around a ferrite core. For the $i$-th nonlinear inductor an often used equation for inductance without hysteresis is\cite{33,34,35n}
\begin{eqnarray}
I_i  = \frac{1}{{L_0 }}\Lambda _i  + \frac{\beta }{{L_0 n^3 }}\Lambda _i ^3,
\label{eq:1}
\end{eqnarray}
\noindent where  $L_0$ is the linear inductance, and the total flux $\Lambda =n\Phi $  is the number of turns times the flux through one turn. The site number $i$ varies from $-p/2+1$ to $p/2$ for $p$ lattice points, and $i=0$ is the center of the lattice. In the last term $\beta $  is the nonlinear parameter and the flux tends to saturate with increasing current $I_i$. The electromotive force across the $i$-th inductor is 
\begin{eqnarray}
V_i (t) = \frac{{d\Lambda _i }}{{dt}} = \frac{{d\Lambda _i }}{{dI_i}}\frac{{dI_i }}{{dt}} = L_s (I_i )\frac{{dI_i }}{{dt}}.
\label{eq:2}
\end{eqnarray}
\noindent According to Eq.~(\ref{eq:1}) the nonlinear inductance, 
\begin{eqnarray}
L_s \left( {\Lambda _i } \right) = \frac{{L_0 }}{{\left( {1 + \frac{{3\beta }}{{n^3 }}\Lambda _i ^2 } \right)}} , 
\label{eq:n3}
\end{eqnarray}
decreases with increasing flux (or current). Since we end up focusing on the flux equation the current expression is given in the appendix. Applying Kirchhoff's law to Fig.~\ref{fig:1}(a) produces the starting equation
\begin{eqnarray}
\dot \Lambda _i  = L_s \left( {I_i } \right)\dot I_i  =  - \frac{{Q_i }}{C} + \frac{{Q_{i - 1} }}{C},
\label{eq:3}
\end{eqnarray}
\noindent where the dot now identifies the derivative with respect to time. The dynamical equation of interest is
\begin{eqnarray}
\ddot \Lambda _i &=& \frac{d}{{dt}}\left[{L_s (I_i )\dot I_i } \right] =\frac{d}{{dt}}\left[\frac{L_0}{1+\frac{3 \beta}{n^3}\Lambda _i^2(I_i)} \dot I_i  \right] \nonumber \\ 
&=& -\frac{1}{C}(I_i - I_{i + 1}) + \frac{1}{C}(I_{i - 1} - I_i), 
\label{eq:4}
\end{eqnarray}
\noindent where $L_s(I_i)$  is the electrical analogue of a nonlinear mass in an inertial lattice. Given the complex saturable nonlinear structure of the current equation it is useful to transform Eq.~(\ref{eq:4}) to a flux equation using Eq.~(\ref{eq:1}). This has the following form:
\begin{eqnarray}
\ddot \Lambda _i  =  - \frac{{\omega _m^2 }}{4}\left( {2\Lambda _i  - \Lambda _{i + 1}  - \Lambda _{i - 1} } \right) \nonumber \\ 
- \frac{{\beta \omega _m^2 }}{{4n^3 }}\left( {2\Lambda _i ^3  - \Lambda _{i + 1}^3  - \Lambda _{i - 1} ^3 } \right),
\label{eq:5}
\end{eqnarray}
where $\omega _m^2  = {4 \mathord{\left/
 {\vphantom {4 {(L_0 C)}}} \right.
 \kern-\nulldelimiterspace} {(L_0 C)}}$  identifies the top of the linear plane wave spectrum and the terms on the far right are analogous to nonlinear onsite potential terms.

\begin{figure}
\includegraphics{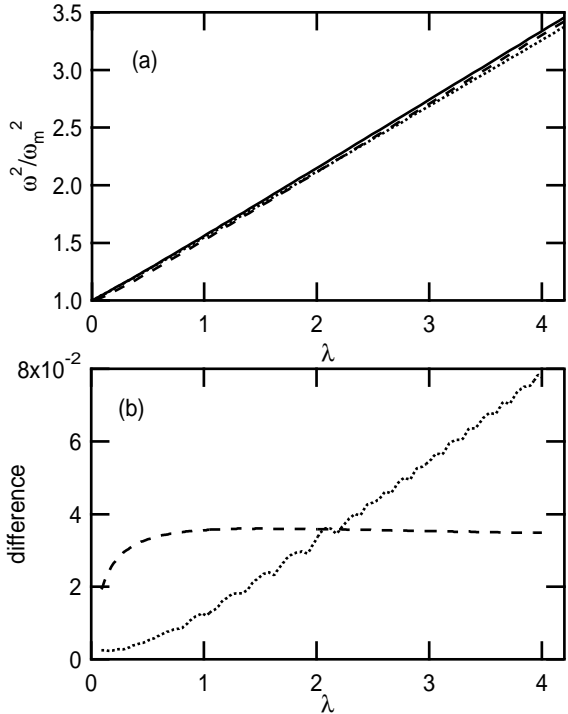}%
\caption{\label{fig:2}
(a) Frequency squared as a function of the nonlinear amplitude parameter  $\lambda $. Solid curve: solution to the eigenvector equation; dashed curve: solution to the three analytical equation approximation; and dotted curve: driven-damped simulations.
(b) Differences between the three kinds of solutions. The solid eigenvector curve is used as a baseline. The other two curves are measured with respect to the solid curve. Dashed curve, the three equation method; dotted is the driven damped method. 
}
\end{figure}

To find the approximate analytical frequency dependence of the flux ILM of odd symmetry as a function of the flux amplitude we follow Ref.~[\onlinecite{35}]. Let
\begin{eqnarray}
\Lambda _i  = \xi _i \cos \omega t,
\label{eq:6}
\end{eqnarray}
\noindent where the center site is 
\begin{eqnarray}
\Lambda _0  = \xi _0 \cos \omega t \equiv  \alpha \cos \omega t, 
\label{eq:7}
\end{eqnarray}
and
\begin{eqnarray}
\xi _i =\left( {-1} \right)^i \alpha Ne^{ - \left| i \right|q'a} = \left( { - 1} \right)^i \alpha N\left( {\frac{1}{y}} \right)^{\left| i \right|}
\label{eq:8}
\end{eqnarray}
\noindent for $\left| i \right| > 0$. Here $q'$ is the imaginary part of the wavenumber, $a$ the lattice constant and the amplitude drops off as $e^{ - \left| i \right|q'a}$  away from the center, with a distinct amplitude ratio, $N/y$, between sites $i=0$  and  $i =  \pm 1$. The center of the odd mode is at  $i=0$ so  $\xi _0  = \alpha $, $\xi _i  = \xi _{-i} $. (The construction of the even symmetry mode is similar and will not be treated here.) Substituting Eq.~(\ref{eq:6}) into Eq.~(\ref{eq:5}) and applying the rotating wave approximation gives a relation for the mode frequency. For the $i=0$  site the result is 
\begin{eqnarray}
\frac{{4\omega ^2 }}{{\omega _m^2 }} = \left[ {2 +2 \frac{N}{y}} \right] +  \lambda \left[ {2 +2 \frac{N^3}{y^3}} \right],
\label{eq:9}
\end{eqnarray}
where the dimensionless nonlinear parameter $\lambda  = \frac{{3\beta \alpha ^2 }}{{4n^3 }}$  depends on the amplitude squared. For the $i=1$  site the appropriate expression is
\begin{eqnarray}
\frac{{4\omega ^2 }}{{\omega _m^2 }} = \left[ {2 + \frac{1}{y} + \frac{y}{N}} \right] + \lambda \left( {2 \frac{{N}^2}{{y^2 }} +  \frac{{N}^2}{{y^5 }} + \frac{y}{N}} \right).
\label{eq:10}
\end{eqnarray}
To estimate the mode frequency and nearest neighbor amplitude we use the condition that any local mode far from its center must obey the general relation\cite{36}
\begin{eqnarray}
4\left( {\frac{{\omega ^2 }}{{\omega _m ^2 }}} \right) &=& \left[ {2 + 2\cosh \left( {q'a} \right)} \right] = \frac{{\left( {y + 1} \right)^2 }}{y}.
\label{eq:11}
\end{eqnarray}
Solving Eqs.~(\ref{eq:9}), (\ref{eq:10}) and (\ref{eq:11}) for $\omega ^2/\omega _m ^2$ , the nearest neighbor amplitude, $\xi _1  =  - \alpha N/y$  and $y$ as a function of $\lambda $  gives the characteristic ILM properties. The frequency dependent results are described by the dashed curve in Fig.~\ref{fig:2}(a), which illustrates that the ILM frequency varies linearly with amplitude, $\alpha $.

\subsection{Driven damped lattice model}
Since there is no general analytic solution for an ILM in this physical lattice we need another procedure to generate  $\Lambda _i(t)$ to compare with the dashed curve shown in Fig.~\ref{fig:2}(a). The next approach is to set up a driven+weak damping arrangement for 50 lattice elements shown in Fig.~\ref{fig:1}(b) and described by
\begin{eqnarray}
 \ddot \Lambda _i  &+& \frac{{\omega _m^2 }}{4}\left( {2\Lambda _i  - \Lambda _{i + 1}  - \Lambda _{i - 1} } \right) \nonumber \\ 
&+& \frac{{\omega _m^2 \beta }}{{4n^3 }}\left( {2\Lambda _i ^3  - \Lambda _{i + 1}^3  - \Lambda _{i - 1} ^3 } \right) + \frac{R}{{L_0 }}\frac{{d\Lambda _i }}{{dt}} \nonumber \\ 
  &=&  - \frac{{C_d }}{{C + C_d }}\left( {\frac{d}{{dt}}V_{d + }  - \frac{d}{{dt}}V_{d - } } \right).
\label{eq:12}
\end{eqnarray}
Here the resistor $R$  provides damping. Since weak damping is to be treated $L_s$  is replaced by $L_0$  in Eq.~(\ref{eq:12}). Parameters for the driving condition are $L_0 /R = 15000/\omega _m $ so that the vibrational life time $\tau  = L_0 /R = 15000/\omega _m $ and a driver strength $2C_d V_{d0} \omega /(C + C_d ) = 3.95 \times 10^{ - 4} n^{3/2} \omega _m ^2 /\sqrt \beta  $,  which is strong enough to move the nonlinear resonance up to $\omega \sim 2\omega _m $. The driving term is $V_{d + }  =  - V_{d - }  = V_{d0} \cos \omega t$.  Starting with a seeded local mode the ILM amplitude is formed and locked to the driver and the seed then removed. The frequency locked ILM amplitude automatically increases the larger the driver frequency difference is from the highest frequency plane wave normal mode. In steady state the time dependent displacement eigenvector  $\Lambda _i(t)$ is obtained for each amplitude. Such simulations show that the ILM is stable. The frequency squared as a function of  $\lambda $ is represented by the dotted line in Fig.~\ref{fig:2}(a). The results are quite close to those found with the approximate analytic three-equation method (dashed curve).

\subsection{Eigenvalue simulations}
To further test these two findings a third method is employed. This is to set the driver-damper $=0$ in Eq.~(\ref{eq:12}) and then solve the equations numerically using Powell's hybrid method with Minpack.\cite{37} Again we assume a time dependence  $\Lambda _i \left( t \right) = \xi _i \cos\omega t$ and apply the rotating wave approximation to the nonlinear terms. This gives a set of eigenvector equations
\begin{eqnarray}
 - \omega ^2 \xi _i  &+& \frac{{\omega _m^2 }}{4}\left[ {2\xi _i  - \xi _{i + 1}  - \xi _{i - 1} } \right] \nonumber \\
&+& \frac{{\lambda \omega _m^2 }}{4}\left( {2\xi _i ^3  - \xi _{i + 1}^3  - \xi _{i - 1} ^3 } \right) = 0,
\label{eq:13}
\end{eqnarray}
with eigenvector solution  $\xi_i$ and frequency $\omega $  at a given amplitude  $\alpha =\xi _0$. One particular ILM solution obtained from the driven-damped simulation is used as an initial condition. The code then finds an ILM eigenvector that satisfies the equations with some tolerance from this initial vector ``guess'' By changing the amplitude slightly from $\alpha $  to $\alpha \pm \Delta \alpha$  other ILM eigenvalues and eigenvectors are generated for these new amplitudes. Continuing this process gives the solid curve in Fig.~\ref{fig:2}(a). Note that the analytic method results are below this curve. Since all three curves are in good agreement the difference between them is plotted in Fig.~\ref{fig:2}(b) using the eigenvector results as the baseline. The three-equation method gives a fixed shift with respect to the solid eigenvector curve while the driven-damped results give better agreement at small amplitude but worse at larger ones.

\begin{figure}
\includegraphics{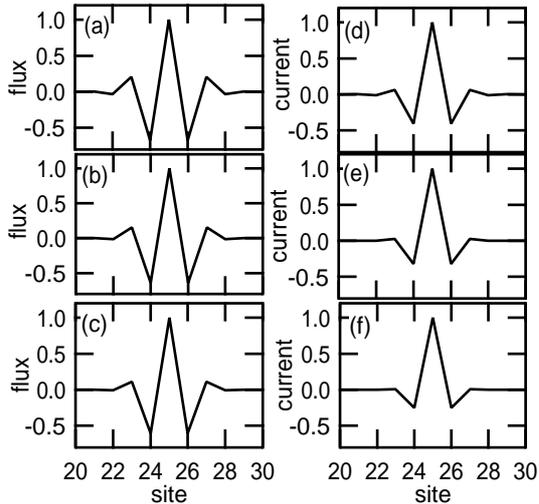}%
\caption{\label{fig:3}
Normalized flux and current eigenvectors for three different amplitude parameters,  $\lambda $. Left column: (a)-(c) shows flux eigenvectors for   $\lambda = 2, 4, 10$ respectively. Right column: (d)-(f) displays current eigenvectors for the same $\lambda $  values. Since each peak amplitude grows with increasing  $\lambda $ each ordinate value is normalized to the amplitude peak.
}
\end{figure}
\section{Discussion}
\begin{figure}
\includegraphics{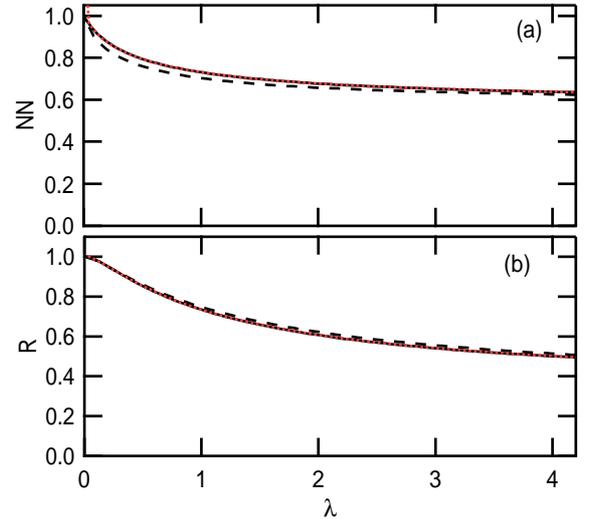}%
\caption{\label{fig:4}
(a) Nearest neighbor amplitude to central element ratio (NN) for the odd flux ILM vs  $\lambda $. Solid curve: eigenvector method; dashed: three equation analytic method; dotted: driven-damped method. At this resolution the solid and dotted curves completely overlap.
(b) Rato $R$ of the current ILM $NN$ to the flux ILM $NN$ vs  $\lambda $. With increasing amplitude the current ILM is spatially compressed to a smaller number of cells compared to the flux ILM. In the asymptotic limit the flux ILM approaches a three-element eigenvector while the current ILM approaches a one-element one.
}
\end{figure}

There is added value in now comparing the flux ILM results with those for the current ILM. Equation~(\ref{eq:1}) is used to make the conversion and a comparison of the eigenvectors for different amplitudes is presented in Fig.~\ref{fig:3}. The left column displays the flux ILM eigenvectors for three different $\lambda $  values while the right column shows the corresponding results for the current ILM. Because the current ILM is spatially compressed to a smaller number of unit cells with respect to the flux ILM its nearest neighbors show a dramatic decrease in relative amplitude with increasing  $\lambda $, indicating that the energy becomes more concentrated in the central cell. A more precise comparison is to plot the nearest neighbor amplitude of the flux ILM divided by the amplitude of the central element as a function of  $\lambda $. We call this ratio NN in Fig.~\ref{fig:4}(a). It is clear that in the asymptotic limit this ratio approaches 0.5, as has been shown earlier to occur for a lattice dynamics chain with hard quartic potential.\cite{38} In Fig.~\ref{fig:4}(b) the ratio R of NN for the current ILM to NN for the flux ILM demonstrates that for the asymptotic limit the current ILM approaches that of a single element excitation.

\section{Summary and Conclusions}
We have demonstrated that an electric transmission line with nonlinear inductors and linear capacitors can give rise to ILMs above the top of the plane wave spectrum. The nonlinear inductor behaves as an onsite nonlinear component, and when the array is transformed to a flux nonlinear transmission line the resulting nonlinear contribution appears as the analogue of an onsite potential. The resulting ILM is relatively straightforward to identify. The flux ILM has been calculated in three different ways: they are the three equation approximate analytic method, a driven damped method in a 50-element lattice and a numerical eigenvalue method for the same lattice. All three methods are in good agreement and show that a current ILM is spatially compressed with respect to the corresponding flux ILM.

To date all nonlinear lattice dynamic studies of inertial systems have focused on the nonlinear potential to produce vibrational ILMs, which, typically, have localized eigenvectors very similar to those of force constant defects in a harmonic lattice\cite{9}. Efforts in a related physics field\cite{14} suggested to us that for inertial lattices with strongly nonseparable, nonlinear, vibrational modes, amplitude dependent masses will need to be considered. To approach this nonlinear lattice problem indirectly we have made use of the well-known lumped element transfer between 1-D electrical and mechanical transmission lines to make use of a nonlinear electrical inductance to understand the dynamical properties of an amplitude dependent inertial mass. Our current study of a monotonic electrical transmission line with an onsite nonlinear inductance indicates that amplitude dependent masses of either nonlinear sign, in a diatomic lattice, should give rise to localized vibrational modes outside of the plane wave spectra. In the large amplitude limit it is expected that they should have eigenvectors very similar to those associated with mass defects in harmonic lattices\cite{39}. Our findings imply the dynamical picture for a strongly, nonseparable, nonlinear lattice will be to replace the system with ILMs plus renormalized phonons. The ILM eigenvectors will be similar to the mass defect and force constant defect types. This nonlinear inductive ILM study strengthens the analogy between the dynamics of defects in the harmonic lattice with ILMs in the strongly anharmonic lattice.

\begin{acknowledgments}
M. S. was supported by JSPS-Grant-in-Aid for Scientific Research No. 25400394. A. J. S. was supported by Grant NSF-DMR-0906491 and he acknowledges the hospitality of the Department of Physics and Astronomy, University of Denver, where some of this work was completed.
\end{acknowledgments}

\appendix*
\section{}
\setcounter{equation}{0}
Finding the current dependence of the nonlinear inductance associated with Eq.~(\ref{eq:n3}) involves solving a cubic equation. We use Cardano's method\cite{40} for the following equation:

\begin{eqnarray}
 t^3  + pt + q = 0
\label{eq:A1}
\end{eqnarray}
where for Eq.~(\ref{eq:1})  $p = \frac{{n^3 }}{\beta }$ and $q =  - \frac{{n^3 }}{\beta }L_0 I$. After some algebra we find
\begin{eqnarray}
\Lambda  &=& \sqrt {\frac{{n^3 }}{{3\beta }}} \times \nonumber \\
\samepage
& &\left( {\sqrt[3]{{J + \sqrt {J^2  + 1} }} + \sqrt[3]{{J - \sqrt {J^2  + 1} }}} \right)
\label{eq:A2}
\end{eqnarray}
where the normalized current is
\begin{eqnarray}
J \equiv \frac{{3^{{3 \mathord{\left/
 {\vphantom {3 2}} \right.
 \kern-\nulldelimiterspace} 2}} \beta ^{{1 \mathord{\left/
 {\vphantom {1 2}} \right.
 \kern-\nulldelimiterspace} 2}} }}{{2n^{{3 \mathord{\left/
 {\vphantom {3 2}} \right.
 \kern-\nulldelimiterspace} 2}} }}L_0 I
\label{eq:A3}
\end{eqnarray}
and
\begin{eqnarray}
\lambda  = \frac{1}{4}\left( {\sqrt[3]{{J_0  + \sqrt {J_0 ^2  + 1} }} + \sqrt[3]{{J_0  - \sqrt {J_0 ^2  + 1} }}} \right)^2 
\label{eq:A4}
\end{eqnarray}
where $J_0$  is the maximum amplitude.

\begin{figure}
\includegraphics{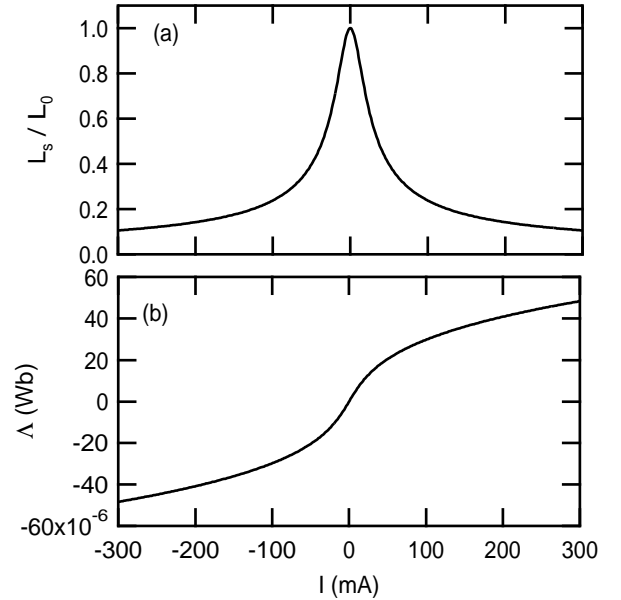}%
\caption{\label{fig:A1}
(a) The current dependence of the nonlinear inductance calculated in Appendix A. (b) The total flux as a function of the current. Half point of the inductance is at 36mA, and the nonlinear parameter at that current is  $\lambda  = 0.25$, while  $\lambda  = 2.11$ at 300mA. Coil dimensions are listed in the appendix.
}
\end{figure}

	A typical current dependence of the nonlinear inductance calculated using Eq.~(\ref{eq:n3}) is shown in frame (a) of Fig.~\ref{fig:A1}. For the linear inductance we assumed a toroidal core made from a Mn-Zn ferrite known as ``75 material".\cite{41} Core dimensions are 12.7mm outer diameter, 7.15mm inner diameter, and 4.9 mm thick. With an effective magnetic pass length $\ell =29.5$mm  and cross section area $s = 1.26 \times 10^{ - 5}$ m$^2$, a 10 turn winding ($n=10$) gives  $L_0  = 615\mu $H by using  $L_0  = \mu _{lin} sn^2 /\ell $. The nonlinear parameter is estimated as follows. The magnetic field $H$ is calculated multiplying Eq.~(\ref{eq:1}) by  $n/\ell $ so 
\begin{eqnarray}
 H &=& \frac{{nI}}{\ell } = \frac{1}{{L_0 }}\frac{n}{\ell }nsB + \frac{n}{\ell }\frac{\beta }{{L_0 n^3 }}n^3 s^3 B^3  \nonumber \\ 
  &=& \frac{1}{{\mu _{lin} }}B + \frac{{\beta s^2 }}{{\mu _{lin} n}}B^3  
\label{eq:A5}
\end{eqnarray}
From the $B-H$ curve of the material and Eq.~(\ref{eq:A5}), the linear and nonlinear parameters are estimated to be $\mu _{lin}  = 13600\mu _0 $  and  $\beta  = 1.205 \times 10^{12} $(1/Wb$^2$), where $\mu _0$  is the magnetic permeability of vacuum. ``75 material" is known as a low-loss material with a small hysteresis. We used the average value of the hysteresis loop to compare with Eq.~(\ref{eq:A5}), over the middle magnetic field region $<$0.35T, smaller than saturation field of 0.43T. The resulting inductance shown in Fig.~\ref{fig:A1} is very nonlinear. According to Eq.~(\ref{eq:A4}) for $I = 300$mA  $\lambda  \approx 2.11$. For completeness frame (b) of Fig.~\ref{fig:A1} presents the dependence of the flux on the current.


\providecommand{\noopsort}[1]{}\providecommand{\singleletter}[1]{#1}%

\end{document}